\documentclass[aps,amsmath,twocolumn]{revtex4}
\usepackage{graphicx}
\usepackage{bm}
\begin{document}
\flushbottom

\title{
Bundling up carbon nanotubes through Wigner defects}

\author{
Ant\^onio J. R. da Silva$^1$, A. Fazzio$^1$ and Alex Antonelli$^2$}

\affiliation{
$^1$Instituto de F\'\i sica, Universidade de S\~ao Paulo, CP 66318,
05315-970, S\~ao Paulo, SP, Brazil\\
$^2$Instituto de F\'\i sica ``Gleb Wataghin'', UNICAMP, CP 6165, 13083-970,
Campinas, SP, Brazil
}

\date{\today}


\begin{abstract}
We show, using {\it ab initio} total energy density functional theory, that the so-called Wigner defects,
an interstitial carbon atom right besides a vacancy, which are present in irradiated graphite can also exist in
bundles of carbon nanotubes. Due to the geometrical structure of a nanotube, however, this defect has a rather
low formation energy, lower than the vacancy itself, suggesting that it may be one of the most important defects that are
created after electron or ion irradiation. Moreover, they form a strong link between the nanotubes in bundles, increasing
their shear modulus by a sizeable amount, clearly indicating its importance for the mechanical properties of nanotube bundles.
\end{abstract}


\maketitle
Single-walled carbon nanotubes (SWNT) have attracted an enormous amount of attention since its discovery\cite{swnt}
about ten years ago. No doubt this is related to all their extraordinaire materials properties and potential for
applications\cite{dress}. From a mechanical point of view, SWNT have exceedingly large elastic modulus\cite{mod}
around 1 TPa, which would make them perfect for low-density high-modulus fibers. When in bundles, however, a
much worse performance is usually obtained, mostly related to the weak interaction between tubes, which makes it relatively easy for them to slide against each other\cite{ajayan}. A breakthrough has been recently obtained\cite{kis} using electron-beam
irradiation to generate crosslinks between the tubes. This increased the shear modulus by a factor of approximately 30
thus indicating a new way to fabricate stronger fibers.

Irradiation by electrons or ions is being used by many groups\cite{irrad}
as an important tool to modify the structure
and properties of nanotubes. As examples, we can mention the welding of tubes\cite{terrones}, and the dramatic
increase of shear modulus for a bundle of nanotubes\cite{kis}. In order to have a better control of the nanotube engineering,
it is fundamental to have a detailed microscopic understanding of the defects that are created upon irradiation. Properties like
formation energy, geometry, stability and electronic structure are critical in determining what type of defects will be produced
and how they will affect the mechanical behavior of the carbon fiber. Some theoretical works have investigated
the effects of electron or ion irradiation on SWNT and SWNT bundles\cite{krash,krash2}.
These studies consisted of Molecular Dynamics simulations
with the interaction between the atoms being described by empirical potentials. Even though these works are relevant and
indicative of the processes that may occur, they may miss important aspects since empirical potentials cannot describe
appropriately features such as re-hybridization, rebonding or charge redistribution, all of them quite important when chemical
bonds are being broken and made, specially in stressed configurations that may result from the irradiation procedure.
Moreover, recombination barriers and formation energies are usually poorly described by these potentials. Yet another
difficulty with these potentials is the important fact that carbon atoms have many possible hybridizations. Therefore, even
though cross links between tubes in bundles have already been predicted to appear after irradiation\cite{krash2}, their detailed
structure and properties that will control how they affect the mechanical behavior of bundles have not yet been determined.
It is thus clear that a fully quantum mechanical description of the electronic structure of the
system is necessary, and that is precisely what we describe below.

All our results are based on {\it ab initio} total energy density functional theory\cite{dft} calculations. We have used
ultrasoft pseudopotentials\cite{pseudo}, a plane wave expansion up to 230 eV, and the local density approximation for the
exchange-correlation potential as implemented in the VASP code\cite{vasp}. A defect-in-supercell approach was employed,
similar to what has been used in the investigation of defects in graphite\cite{tell,ewels}.
We have in our supercell two (5,5) nanotubes, with 100 atoms each, with lattice primitive vectors appropriate
for a hexagonal arrangement of tubes in the bundle, as shown in Fig. 1(a). The Brillouin zone sampling had a total of
six points, with three of them along the tube axis direction. In all calculations the positions of the unconstrained
atoms in the supercell were relaxed until the forces in them were smaller than 0.02 eV/$\text{\AA}$. We obtain
1.42 $\text{\AA}$ for the C-C bonds perpendicular to the tube axis, and 1.41 $\text{\AA}$ for the others, which gives
for the cell length the value of 12.21 $\text{\AA}$. For the inter-tube distance in the bundle we obtain approximately
a center-to-center distance of 9.87 $\text{\AA}$. From a variety of tests we estimate that the errors in energy differences
due to all our approximations are of the order of 0.05 eV.

\begin{figure}[ht]
\includegraphics[angle=0,width=7.cm]{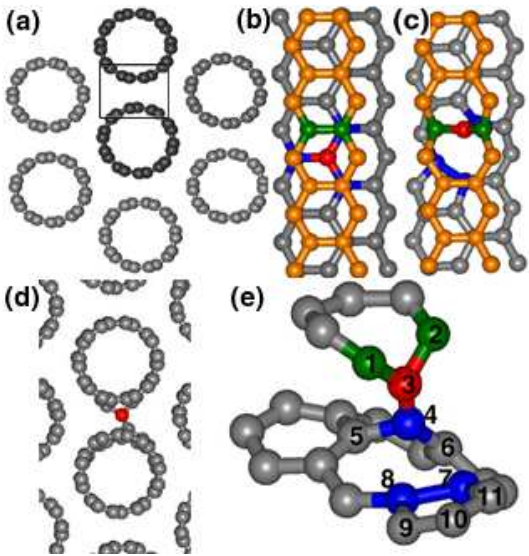}
\caption{
(a) Final geometry for a SWNT bundle, showing the hexagonal order and the unit cell (tubes in black).
(b) Detail of the ``line-of-contact'' between the tubes in the defect free
configuration (region marked by the rectangle in (a)), as seen from the inside of the top tube.
The orange (grey) atoms are in the top (bottom) tube. The atom that will form a link between the two tubes is marked
in red, whereas its three nearest neighbours are marked in blue. Upon displacement, the red atom will form bonds to the
two atoms in the top tube marked in green. (c) Same atoms after the formation of the defect.
(d) Final geometry for the configuration with the Wigner defect. The interstitial atom (red) can be clearly
seen connecting two tubes. (e) Detailed view of the Wigner defect. The pentagon formed after the
interstitial is ejected can be clearly seen (atoms 7 through 11).}
\end{figure}

The two tubes in the unit cell are arranged in such a way as if they were obtained by rolling up two neighboring graphene
sheets in a graphite structure, in other words, at the ``line-of-contact'' between them the atomic arrangement resembles that
of graphite (see Fig. 1(b)). Since the inter-tube interaction is weak, they have a lot of rotational freedom
within the bundle at room temperature\cite{kwon}, which does not make this choice restrictive. However, the curvature will
play an important role in the stability of the defect, as discussed below. The structures of a variety of defects in graphite
have been recently studied\cite{tell,ewels}. In particular, an intimate Frenkel pair defect has been suggested to be responsible
for the undesirable energy release in graphite moderators in nuclear reactors, the Wigner energy. The procedure we
employed in the attempt to produce a Frenkel pair similar to what was found in graphite\cite{tell,ewels}, was the following:
we moved an atom in one of the tubes from the ``line-of-contact'' and placed it approximately in between the two tubes,
in the middle of a close by C-C bond in the other tube. The C atoms in the farthest away planes (in both tubes, 20 atoms total)
from the defect were held fixed throughout the atomic relaxations. This was done in order to simulate an isolated
defect, where the atoms far from the perturbation would not move. During the first few relaxation steps the displaced
atom was also kept fixed, otherwise it would tend to return to its original position. All the attempts to create a
Frenkel pair defect by displacing one atom from the top tube of Fig. 1(b) (one of the orange atoms) were unsuccessful.
Due to the curvature of the tube, the atom always returned to its original position.  Eventually, tube rotations may
be able to stabilize such configurations, but this may depend on the temperature and the time scales for rotation
as compared to the defect relaxation time. On the other hand, when we displaced an atom from the bottom tube
of Fig. 1(b), a stable Frenkel pair was obtained since the ``line-of-contact'' and curvature were both favorable.

\begin{figure}[ht]
\includegraphics[angle=0,width=7.cm]{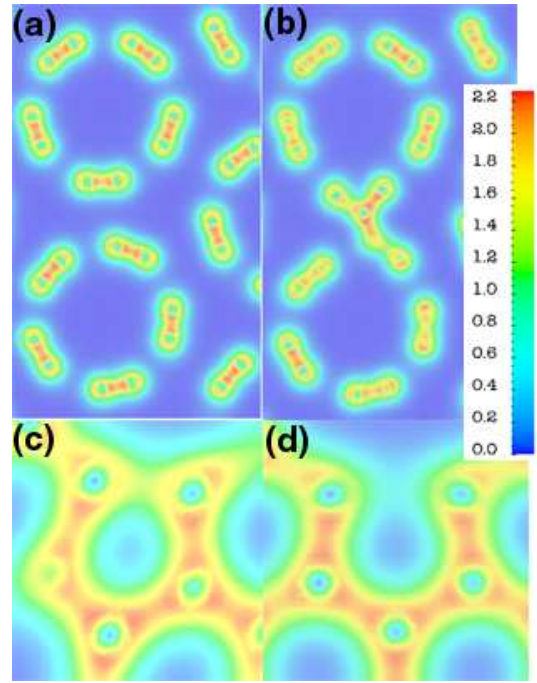}
\caption{
Charge density (e$^-$/$\text{\AA}^3$) in: (a) plane perpendicular to
the tubes axis and farther away from the Wigner defect. Five C-C bonds in each tube can be clearly seen. (b) Plane
passing through atoms 1-2-3-4 in Fig. 1(e). The C-C bond that existed between atoms 1 and 2 in the upper tube is broken,
and two new bonds are formed with the interstitial atom.
(c) Plane that approximately passes through the five atoms marked in  Fig. 1(e) from 7 to 11.
(d) Similar to (c) in graphite.}
\end{figure}

As shown in Fig. 1, the displaced atom (marked in red and labeled 3 in Fig. 1(e)) forms
two bonds with C atoms of the top tube (marked in green, and labeled 1 and 2 in Fig. 1(e)). The bond length between
atoms 1 and 2 increases from their equilibrium value of 1.42 $\text{\AA}$ to 2.16 $\text{\AA}$, which clearly indicates
that it is broken. This can also be seen in Fig. 2 via the charge density plots. On the other hand, the distances between
carbons 1-3 and 2-3 (Table I) clearly show the formation of two C-C bonds. Curvature effects and the constraints being
imposed lead to different bond lengths. The displaced
atom remains bonded to its original tube through one of its original neighbors, the atom labeled 4 in Fig. 1(e).
The distance 3-4 decreases from 1.41 $\text{\AA}$ to 1.36 $\text{\AA}$ upon the formation of the defect. This is
suggestive of a sp$^2$ hybridization for these two carbon atoms with the formation of a double bond between them,
since the C-C distance in the H$_2$C=CH$_2$ molecule is 1.34 $\text{\AA}$. These results indicate that a strong bond
is being formed between the two tubes, as can be seen from the charge density plots of Fig. 2. The atom 4 remains
linked to its two other neighbors via the bonds 4-5 and 4-6, with bond lengths diplayed in Table I.
As a reference for comparison, we have also studied the
Wigner defect in graphite\cite{tell,ewels}. We have used a similar approximation as described above, however the
supercell now had 100 C atoms, 50 in each of the graphite planes, and 100 k-points were used for the Brillouin zone
sampling. The equilibrium C-C bond length was 1.41 $\text{\AA}$ and the interplanar distance was 3.30 $\text{\AA}$.
Upon the formation of the Frenkel pair, we obtained the equivalent distances shown in Table I, which
indicate that the links caused by the defects in the nanotube bundles will most likely be stronger than in graphite.
\begin{table}
\caption{Interatomic distances ($\text{\AA}$) for the atoms labelled 1 through 11 in Fig. 1(e).
Results are presented for the SWNT bundle when some of the atoms were fixed (line I),
when they were all allowed to relax (line II), and for graphite (line III).}
\label{tab1}
\begin{tabular}{cccccccccccc}
 & 1-2 & 1-3 & 2-3 & 3-4 & 4-5 & 4-6 & 7-8 & 8-9 & 9-10 & 10-11 & 7-11\\
\hline
I & 2.16 & 1.50 & 1.41 & 1.36 & 1.42 & 1.50 & 1.57 & 1.41 & 1.41 & 1.40 & 1.42\\
II & 2.21 & 1.45 & 1.43 & 1.33 & 1.44 & 1.45 & 1.54 & 1.42 & 1.40 & 1.40 & 1.43\\
III & 1.57 & 1.45 & 1.44 & 1.32 & 1.44 & 1.44 & 2.00 & 1.39 & 1.40 & 1.40 & 1.39\\
\end{tabular}
\end{table}

An important aspect of this defect in nanotube bundles when compared to graphite is the atomic relaxation around the
defect. When the C atom is displaced, it remains bound to one of its three former nearest neighbors (the atoms
marked in blue in Fig. 1), as discussed above. The other two C atoms are now in an energetically
unfavorable configuration, and will tend to approach each other in an attempt to make a chemical bond. This will
lead to the formation of a pentagon, which can be clearly seen in Fig. 1(e), where the five C atoms are labeled
from 7 through 11. In a nanotube, however, due to its curvature allied to a certain degree of flexibility towards
compression, these two C atoms can get relatively close to each other, with a final interatomic distance of
1.57 $\text{\AA}$. This should be contrasted to the similar distance in graphite, 2.00 $\text{\AA}$.
Therefore, in the case of nanotubes these two carbon atoms can rebond much more effectively, which should have
an important contribution to the stability of the defect. In Fig. 2(c) and Fig. 2(d) we present charge density
plots in planes that pass through the atoms that form this pentagon, for both the nanotube and graphite. It can be
clearly seen that there is indeed a much stronger rebonding in the nanotubes when compared to graphite. It should
be noted that due to the curvature of the nanotube, it is very difficult to have a plane passing exactly through all five
C atoms (from 7 through 11), that is the reason why atom 9 (see Fig. 1(e)) appears less clear in Fig. 2(c).
All the relevant interatomic distances are summarized in Table I.

We calculate the defect formation energy as the difference between the total energy for the system with and without
the defect, which gives a value of 7.27 eV. For the Wigner defect in graphite we obtain a formation energy of 10.67 eV,
similarly to other authors\cite{tell,ewels}. This means that it is much easier to produce these defects in nanotube bundles
than in graphite, indicating that it may be very common in irradiated samples. If we consider a vacancy in an isolated
nanotube, we obtain for a fully relaxed structure a formation energy of 5.82 eV. However, if we now also allow all the
atoms in the bundle to relax, i.e., the 20 C atoms that were held fixed can now relax, we obtain a formation energy of
5.53 eV. This is an extremely low value, showing that this defect may be one of the most important defects in nanotube
bundles. The full relaxation allows the tubes to orient themselves in such a way that their interlink bonds are stronger,
as indicated by the bond lengths shown in Table 1. Therefore, if the tubes have enough mobility to accommodate the
defect, or if the defect concentration is high enough, we expect a formation energy close to 5.5 eV. However, if the
defect is isolated and the tubes cannot freely move the formation energy may be closer to 7.3 eV.

\begin{figure}[ht]
\includegraphics[angle=0,width=7.cm]{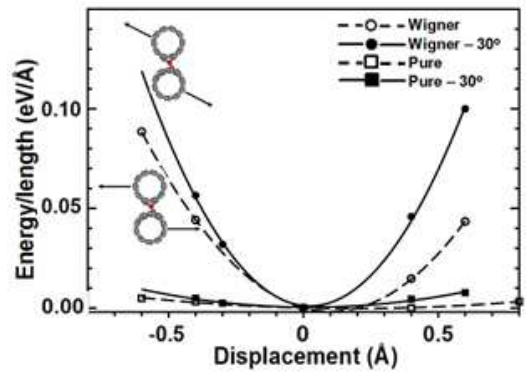}
\caption{Change in energy (per unit length) caused by the relative displacement of two isolated tubes (not in a bundle).}
\end{figure}

The strong link that is being formed between the tubes may have a dramatic change in the mechanical properties
of nanotube bundles. To estimate the changes in the shear modulus, we perform calculations for two isolated tubes,
instead of a full bundle. This should give the correct trend, since the tube-tube interaction is much stronger in the
presence of the Wigner defect. We perform relative displacements as illustrated in Fig. 3. For the displacements
perpendicular to an imaginary line that joins the tubes axes we obtain the dashed curves, whereas for
displacements making a 30$^o$ angle with respect to the previous displacement direction we have the solid curves.
This latter direction would be more representative of a shear displacement of tubes in a bundle. For the
dashed curves we obtain shear moduli of 3 GPa and 59 GPa for the pure and defected tubes, respectively, whereas
for the solid curves we obtain 7 GPa and 97 GPa, respectively. This indicates an increase in the shear modulus by a
factor of 20 in the former case and by a factor of 14 in the latter case. Therefore, we show that this defect may be the
main reason behind the observed increase in shear modulus of nanotube bundles upon irradiation\cite{kis}. In the work
by Kis and collaborators\cite{kis} it has been proposed that carbon interstitials or radiation induced chemical reactions
involving, for example, carboxyl groups could form links between the nanotubes in a bundle. These changes would
be responsible for the increase in shear modulus, however, no estimates for the formation energies of these structures
are provided. We here propose a similar idea but with an alternative local atomic structure, which due to its low formation
energy is very likely to be created during the irradiation.

\begin{figure}[ht]
\includegraphics[angle=0,width=7.cm]{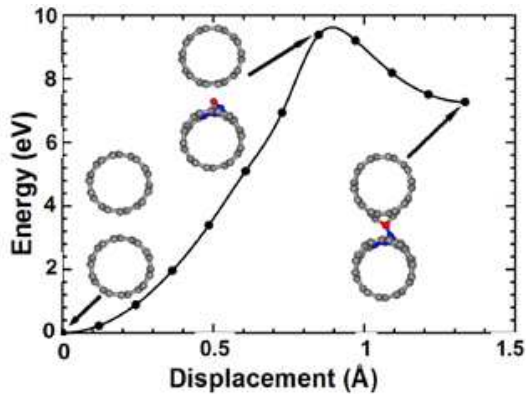}
\caption{
Energy as a function of displacement of the interstitial atom from its ideal configuration (zero displacement).}
\end{figure}

Finally, an important question is related to the stability of such a defect. In graphite, it has been proposed\cite{tell}
that this defect is responsible for the Wigner energy release peak at 200$^o$C, associated with an experimental
barrier\cite{mitchel} of approximately 1.4 $\pm$ 0.4 eV, which agrees nicely with the calculated barrier\cite{tell,ewels}
of 1.3 eV for the defect recombination. To estimate the barrier, we linearly interpolate the coordinates of the displaced
atom between its position in the pure bundle (Fig. 1(a)) and in the final defect configuration (Fig. 1(d)). We then held this
atom fixed and let all the other atoms relax, except the 20 constrained atoms as discussed above. Ten intermediate
configurations were calculated, and the results are plotted in Fig. 4. We obtain a recombination barrier of
approximately 2.4 eV, indicating that the defect is significantly more stable in nanotube bundles than in graphite.
If we assume that the recombination of this defect would also be associated with an energy release, like in graphite,
and that to observe such an energy release it would require a recombination rate of the same order than that in graphite,
we estimate that in nanotube bundles such a release should be observed around 600$^o$C.

In summary, we have identified a very stable Frenkel pair in nanotube bundles, able to provide strong
links between nanotubes, thus altering in a dramatic way the mechanical properties of these bundles. Since these defects
have rather low formation energies, they should be prevalent in irradiated samples. They should play an important role in
nanotube engineering using irradiation sources, a field that has been gaining a lot of attention lately. Besides forming the
strong links, as already mentioned, they could be spots where welding of two tubes will start. Moreover, since they provide a
somewhat open local structure in one of the tubes (a small ``hole'' in the tube wall), they may allow the entrance of atoms or
small molecules inside the tube, which may be important for storage purposes\cite{meunier}. Even though we focused on
a (5,5) nanotube, the conclusions are not dependent on its chirality. Finally, it has not escaped our attention that a similar
structure will probably exist in multi-walled nanotubes linking, for example, two neighboring walls\cite{huhtala}. Some of the
ideas discussed here were already foreseen in the study of defects in graphite\cite{tell}, however, the structure of nanotubes
is such that the defect has a combined high stability and low formation energy, which could not be expected from
graphite studies.

We acknowledge support from the Brazilian Agencies FAPESP and CNPq.

\end{document}